\documentclass[prb,twocolumn,showpacs,preprintnumbers,amsmath,amssymb]{revtex4}

\usepackage{graphicx}
\usepackage{dcolumn}
\usepackage{bm}

\begin{document}

\title{Exactly solvable model of electron in the  Lame potential \\
and singularities of the electron thermodynamic potential}

\author{Victor~G.~Baryakhtar}
\email{bar@imag.kiev.ua}
\author{Eugene~D.~Belokolos}
\email{bel@imag.kiev.ua}
\author{Oleksandr~V.~Dmytriiev}
\email{dmitr@univ.kiev.ua} \affiliation{Institute of Magnetism,
Vernadski Blvd. 36-b, 03142 Kyiv, Ukraine}

\begin{abstract}
One-gap and two-gap separable Lame potentials are studied in detail.
For the one-dimensional case, we construct the dispersion relation
graph $E(k)$ and for the three-dimensional case we construct the
Fermi surfaces in the first and second bands. The pictures
illustrate a passage from the limit case of free electrons to the
limit case of tight binding electrons. These results are used to
describe the Lifshits electron phase transition of $2 \frac{1}{2}$
kind and derive some exact expressions. We also examine the
singularities of the second derivative of magnetic momentum in an
external magnetic field. The parameter of the singularities depends
on corresponding effective mass.
\end{abstract}

\maketitle

\section{Introduction}
There is a sufficient number of numerical and self-consistent
methods to compute the one particle spectrum of a solid with good
accuracy. However, in addition to this, it is desirable to have
analytical  solutions of the Schr\"{o}dinger equation. These
solutions for multidimensional periodic potentials are useful both
for  a zero order  perturbation theory and for the general theory.
This allows us, for example, to gain a deeper understanding of
spectral manifolds and corresponding eigenfunctions, as well as
their geometrical and analytical properties.

Among one-dimensional potentials, the finite-gap ones  are the most
interesting and fruitful in the view of the physical results. The
Schr\"{o}dinger equation with the finite-gap potential \cite{ZMNP,
BBEIM} has an exact solution,  the spectrum and eigenfunctions are
defined by analytical functions.

The Kronig-Penny potential \cite{KP} consisting of rectangular wells
is usually considered as the most elementary one in the quantum
physics of solids. However, in order to determine the band structure
that corresponds to this potential, it is necessary to solve
transcendental equation and, furthermore, the appropriate
eigenfunctions are too cumbersome  to use them to calculate the
matrix elements of any observables. The finite-gap potentials are
free of these limitations. The matrix elements of any observable on
the eigenfunctions that correspond to finite-gap potential can be
calculated analytically by the residue method. Furthermore, one can
say that these potentials play the same role in solid-state physics
as the Kepler problem plays in atomic theory.

It is known that generic periodic potential has generally an
infinite set of gaps in the spectrum, but their width rapidly
decreases as energy increases. If a potential is analytical, then
the rate of decrease is exponential. Therefore, provided narrow
enough gaps in the spectrum are disregarded, any periodic potential
can be approximated by a finite-gap potential. \cite{MO} It turns
out also that the finite-gap potentials are exact solutions of the
Peierls-Fr\"{o}hlich problem on a self-consistent state of the
lattice and conduction electrons. \cite{B}

The simplest finite-gap potentials are the Lame ones. \cite{I, I1,
BEES, T} They are expressed through the Weierstrass function. The
Lame potentials are a subset of the Darboux potentials, \cite{D}
that are linear superpositions of the Weierstrass functions with
half-period shifts. Whatever the number of gaps, all these
potentials are characterized by only two parameters of the elliptic
function. Treibich and Verdier rediscovered the Darboux potentials
and described them in algebraic-geometric terms of the tangent
covers and studied their spectral properties. \cite{T1, TV, TV1}
Krichever suggested considering elliptic finite-gap potentials from
the point of view a torus covering. \cite{K} Using this approach,
Smirnov has presented a large list of these potentials. \cite{S} Its
and Matveev derived a general formula that expresses an arbitrary
finite-gap potential through the multidimensional Riemann
$\theta$-functions, the parameter number of which coincides with the
number of band edges. \cite{IM}

There are also multidimensional elliptic Calogero-Moser potentials,
which their integrability was proved by Olshanetsky and Perelomov.
\cite{OP}

Baryakhtar, Belokolos and Korostil suggested considering the
Schr\"{o}dinger equation with three-dimensional separable potential
as a sum of three  one-dimensional finite-gap potentials along
orthogonal directions. \cite{BBK2} It is obvious that these
potentials correspond to the orthorhombic lattice. The
Schr\"{o}dinger equation with such a potential has the exact
eigenfunctions and the spectrum is defined by analytical functions.
The parameters that this model includes have simple physical sense.
These parameters are related to the width of spectral gaps. As an
example, let us consider separable elliptic finite-gap potentials.
The Weierstrass function has two independent parameters: the
parameter $\omega'$ defines the potential period, the parameter
$\omega$ defines the gap width along corresponding direction of dual
lattice. It is natural to consider the parameter $\tau={\omega'} /
{\omega}$ as a characteristic of the relative width of gaps. As the
value of $|\tau|$ increases, the corresponding gap width increases
too. In one of the extreme cases, the width of all the gaps
vanishes. It corresponds to the free electron model, while in the
other extreme case, the width of all the bands vanishes and this
corresponds to the tight binding model. In the model of separable
finite-gap potentials, we can also easily study one- and
two-dimensional lattices. This can be done by choosing extremely
wide gaps along certain directions.

Finite-gap potentials have successfully been applied  to solve
different problems in solid-state physics. \cite{BBK1, BBK3, BBSD,
BES, BP, BP1}

In our paper, on one hand, we  examine the known one-gap and two-gap
separable Lame potentials in detail. For the one-dimensional case,
we construct the dispersion relation graph $E(k)$  and for the
three-dimensional case, we construct the Fermi surfaces in the first
and second bands. The following pictures illustrate a change from
one extreme case of free electrons to the other extreme case of
tight binding electrons. On the other hand, we use the separable
one-gap Lame potential to describe the Lifshits electron phase
transition of $2 \frac{1}{2}$ kind and derive some exact
expressions. At the end, we consider singularities of the second
derivative of the magnetic momentum of a normal metal in an external
magnetic field.

The paper is organized  as follows. In section 2, we present
information on the one-dimensional Lame potentials using two
different approaches and the one-gap and two-gap ones are studied in
detail. In section 3, we consider the separable finite-gap Lame
potentials and construct the Fermi surfaces using the one-gap one.
In section 4, the one-gap separable Lame potential is applied to
describe the Lifshits electron phase transition of $2 \frac{1}{2}$
kind. At the end of this section we consider singularities of the
second derivative of the magnetic momentum of a normal metal in an
external magnetic field.

\section{The one-dimensional finite-gap Lame potentials}
The Lame potentials can be represented in one of the two forms
(since these potentials are expressed through the elliptic functions
which have two independent periods)
\begin{equation}
V(x)=\ell\left(\ell+1\right)\wp\left(x+\omega' | \omega, \omega' \:
\right), \label{1b}
\end{equation}
\begin{equation}
U(x)=-\ell\left(\ell+1\right)\wp\left(i x + \omega | \omega, \omega'
\: \right), \label{2b}
\end{equation}
were $\ell$ is a number of gaps in the spectrum, and $\wp\left(z |
\omega, \omega'\: \right)$ is the elliptic Weierstrass function with
a real half-period $\omega$ and an imaginary one $\omega'$. We use
the standard notation of the elliptic function theory. \cite{BE} If
it doesn't cause misunderstanding we will write simply $\wp\left(z
\right)$.

The potential $V(x)$ has the period $2 \omega$, the period of the
potential $U(x)$ is $2|\omega'|$. Since the Weierstrass function is
homogeneous and it isn't changed under unimodal transformations, the
potentials (\ref{1b}) and (\ref{2b}) are expressed one through the
other: $\ell(\ell+1)\wp(x+\tilde{\omega}'  | \tilde{\omega},
\tilde{\omega}') = - \ell(\ell+1)\wp(ix+ i\tilde{\omega}' |
i\tilde{\omega}, i\tilde{\omega}' \:) = -\ell(\ell+1)\wp(ix+ \omega'
\: | \omega, \omega')$. We put $\omega=|\tilde{\omega}'|$, and
$\omega'=i\tilde{\omega}$; correspondingly $\eta=-|\tilde{\eta}\:'|,
\:\: \eta '=-i \tilde{\eta}$ and $g_{3}=-\tilde{g}_{3}, \:\:
g_{2}=\tilde{g}_{2}$, also $e_{1}=-\tilde{e}_{3},\:
e_{3}=-\tilde{e}_{1},\: e_{2}=-\tilde{e}_{2}$. In order to pass from
the spectrum of the Schr\"{o}dinger operator with the potential
$U(x)$ to  the spectrum of the Schr\"{o}dinger operator with the
potential $V(x)$, it is necessary to pass from $\omega, \: \omega'$
to $\tilde{\omega}, \: \tilde{\omega}' \: $ in all the expressions.
This means that we have to write $-E$ instead of $E$,
${\eta'}/{\omega'}$ instead of ${\eta}/{\omega}$ and $i k$ instead
of $k$ in expressions for the spectrum that will be defined later.

We will examine the potentials $U(x)$ only.

The dimensionaless Schr\"{o}dinger equation \cite{ZMNP}
\begin{equation}
-\partial _{x}^{2} \Psi- \ell\left(\ell+1\right)\wp\left(i x +\omega
  \right) \Psi = E \Psi  \label{3b}
\end{equation}
has two Bloch eigenfunctions which correspond to different signs of
the wavevector
\begin{equation}
\Psi_{+}(x,E)=\sqrt{\Lambda(x,E)}
\exp{\left(\frac{W(E)}{2}\int{\frac{dx}{\Lambda(x,E)}} \right)},
\label{4b}
\end{equation}
\begin{equation}
\Psi_{-}(x,E)=\sqrt{\Lambda(x,E)}
\exp{\left(-\frac{W(E)}{2}\int{\frac{dx}{\Lambda(x,E)}} \right)}.
\label{5b}
\end{equation}
In order to pass to dimensional values it is necessary to multiply
$E$, $U(x)$ by $E_{0}=\hbar^{2} / 2ma^{2}$ and spacial variable $x$
by $a$, were $a$ is a characteristic length of the crystal
potential; $m$ is a particle mass and $\hbar$ is the Planck
constant.

W(E) is the Wronskian of the functions $\Psi_{-}(x,E)$,
$\Psi_{+}(x,E)$, and $W^{2}(E)$ is a polynomial with respect to $E$.
The roots of this polynomial are spectral edges,
\begin{equation}
W^{2}(E)=-4\prod_{j=1}^{2\ell+1}(E-\varepsilon_{j}). \label{6b}
\end{equation}

The Bloch eigenfunction $\Psi_{\pm}$ and wavevector  $k$ are
holomorphic functions with respect to complex variable $E$ and they
are defined on the Riemann surface $\sqrt{-W^2(E)}$. \cite{ZMNP}
This surface  will be denoted by  $\Gamma$.

The $\Lambda(x,E)$ is  a product of the two Bloch eigenfunctions of
the Schr\"{o}dinger equation (\ref{3b}). Since the eigenfunction
$\Psi_{\pm}(x,E) \sim \exp(\pm i\sqrt{E} x)$ at $E \rightarrow
\infty$, it follows that $\Lambda(x,E)$ with respect to the variable
$E$ is a polynomial of the $\ell$-th order,
\begin{align}
\Lambda(x,E)=\Psi_{+}\Psi_{-} = C&\prod_{j=1}^{\ell}(\gamma_{j}(i x
+ \omega)-E) \nonumber  \\  &= \sum^{\ell}_{j=0}C_{j}(E) (\wp(i x
+\omega ))^{\ell-j}. \label{7b}
\end{align}

The roots $\gamma_{n}(i x +\omega)$ are disposed inside gaps or on
their edges and there is only one root for every gap. From the
$\Lambda(x,E)$ definition, it follows that functions $\gamma_{n}(i x
+\omega)$ are periodic with a period $2|\omega'|$. When the variable
$x$ is changed, the function $\gamma_{n}(i x +\omega)$ takes on
values in the $n$-th gap and extreme values of this function, which
are defined by equation $\partial_{x} \gamma_{n}(i x +\omega),$
coincide with the $n$-th gap edges.

In every band the $\Lambda(x,E)$ maintains the sign. When it passes
from one band to another the sign  changes. In the first band the
$\Lambda(x,E)>0, \: \forall x$. It is easy to verify that
$\Lambda(x,E)$  satisfies the third order equation
\begin{align}
\partial_{x}^{3} \Lambda4  [ \ell(\ell+1)\wp & \left(i x +\omega
\right)+E]  \partial_{x}\Lambda \nonumber \\ &+ 2  \ell(\ell+1)
\wp_{x} \left(i x +\omega \right)\Lambda=0.  \label{8b}
\end{align}
Functions $\gamma_{j}(i x +\omega)$ or coefficients $C_{j}(E)$ can
be found if we insert the $\Lambda(x,E)$ into the equation
(\ref{8b}).

If we know function $\Lambda(x,E)$, we can find the band edges by
equation \cite{WW}
\begin{align}
W^{2}(E)=-4 &[E+\ell(\ell+1) \wp(ix+\omega)] \Lambda^{2}(x,E)
 \nonumber \\ &-2\Lambda(x,E)
\partial^{2}_{x}\Lambda(x,E)+(\partial_{x}\Lambda(x,E))^{2}. \label{9b}
\end{align}

The wavevector $k(E)$ is defined by expression
\begin{equation}
k(E)= \frac{W(E)}{4 \omega'
}\int\limits_{x_{0}}^{x_{0}+|2\omega'|}{\frac{dx}{\Lambda(x,E)}}=\frac{W(E)}{2i}
\left< \frac{1}{\Lambda(x,E)} \right> \label{10b},
\end{equation}
where
\begin{equation*}
\left<f\right>=\frac{1}{2|\omega'|}\int\limits_{x_{0}}^{x_{0}+2|\omega'|}
f(x) dx
\end{equation*}
is the averaging with respect to the variable $x$.

When $E$ takes on values in a gap, the function $\Lambda(x,E)$
alternates its sign and the integral (\ref{10b}) exists only in
sense of the principal value.

The function $\Lambda (x,E)$ is periodic with respect to the
variable $x$ and has a period  $|2\omega'|$. It means that
\begin{equation*}
\frac{W(E)}{4 \omega' }\int {\frac{dx}{\Lambda(x,E)}} = \frac{1}{2
\omega' }\left[ F(x,E)+i k(E)x \right]+C,
\end{equation*}
where  $F(x,E)$ is a certain periodic function with respect to the
variable $x$ with a period $|2\omega'|$. This function and the
function under the integral have poles in the same place.

The expression (\ref{10b}) can be transformed into a more convenient
form. We will use  the following designation
\begin{equation*}
\chi(x,E)=\frac{\Lambda(x,E)}{W(E)}.
\end{equation*}
Then, if we pass to the new function $\chi (x,E)$ in the expression
(\ref{9b}) and differentiate it with respect to the variable $E$, we
obtain
\begin{align}
\frac{\partial \chi  ''}{\partial E}=  -2\chi-4 & [\ell(\ell+1)
\wp(i x+\omega)+E]\frac{\partial \chi}{\partial E} \nonumber \\
&-\frac{\chi  ''}{\chi} \cdot \frac{\partial \chi}{\partial
E}+\frac{\chi' }{\chi} \cdot \frac{\partial \chi'}{\partial E}.
\label{11b}
\end{align}
Here and further, we denote a derivative with respect to $x$ by
prime. The function
\begin{equation*}
-\frac{\partial \chi  '}{\partial E}+\frac{\chi' }{\chi} \cdot
\frac{\partial \chi}{\partial E}
\end{equation*}
is periodic with respect to $x$, and
\begin{equation}
\left<  \left( -\frac{\partial \chi '}{\partial E}+\frac{\chi'
}{\chi} \cdot \frac{\partial \chi}{\partial E} \right)^{'} \:
\right>=0. \label{12b}
\end{equation}
Using formulas (\ref{9b}) and (\ref{11b}) we obtain the following
equality,
\begin{equation}
\left( -\frac{\partial \chi '}{\partial E}+\frac{\chi ' }{\chi}
\cdot \frac{\partial \chi}{\partial E} \right)^{'}  =2 \chi +
\frac{\partial}{\partial E}\left( \frac{1}{\chi} \right).
\label{13b}
\end{equation}
Using formulas (\ref{10b}), (\ref{12b}) and (\ref{13b}) we derive
the expression for wavevector
\begin{equation}
k(E)= k_{0}+\int\limits_{E_{0}}^{E}\frac{\left< \Lambda(x, z)
\right> \: dz}{2\sqrt{\prod_{j=1}^{2\ell+1}(z-\varepsilon_{j})}}.
\label{14b}
\end{equation}
The expression (\ref{14b}) defines a dependence of the electron
energy $E$ on the wavevector $k.$ This function is monotonous, but
piecewise analytical, it corresponds to the extended band scheme. In
the contracted band scheme the energy is periodic analytical
function of wavevector in dual lattice  with a period of this
lattice (in our case this period is equal to ${2\pi} /
{2|\omega'|}$).

Let us introduce $\ell$ parameters $\{t_{j}\}$ by means of the
expression
\begin{equation}
\Lambda(x,t_{1},...,t_{j})=\Psi_{+}\Psi_{-}=C\prod_{j=1}^{\ell}(\wp(i
x +\omega)-\wp(t_{j})). \label{17b}
\end{equation}
Then, in terms of these parameters, we can present the functions
$\Psi_{\pm}(x,E)$ and $k(E)$ in another form, \cite{WW}
\begin{widetext}
\begin{align}
\Psi_{+}(x,k)= u(x,k) \exp(ikx)= \prod_{j=1}^{\ell}\left[
\frac{\sigma(t_{j}-i x-\omega)}{\sigma(i x+\omega) \sigma(t_{j})}
\right]\exp{\left(  i x \sum_{j=1}^{\ell}\frac{\eta'}{\omega'}t_{j}
\right)} \exp(ikx), \label{15b}
\end{align}
\begin{align}
\Psi_{-}(x,k)= \bar{u}(x,k) \exp(-ikx)=\prod_{j=1}^{\ell}\left[
\frac{\sigma(t_{j}+i x+\omega)}{\sigma(i x+\omega)\sigma(t_{j})}
\right]\exp{\left(  -i x \sum_{j=1}^{\ell}\frac{\eta'}{\omega'}t_{j}
\right)} \exp(-ikx), \label{16b}
\end{align}
\begin{align}
k(t_{1},...,t_{j})=\sum_{j=1}^{\ell}\left( \zeta(t_{j})
-\frac{\eta'}{\omega'}t_{j} \right). 
\end{align}
\end{widetext}

Inserting the expression  (\ref{17b}) into (\ref{8b}), we derive a
system of the algebraic equation \cite{BEES} to determine the
quantities $\wp(t_{j})$:
\begin{equation}
\alpha _{k} ES_{k - 1} + \beta _{k} g_{2} S_{k - 2} + \gamma _{k}
g_{3} S_{k - 3} - \delta _{k} S_{k} = 0, \quad k = \overline
{1,\ell} \label{19b}
\end{equation}
where
\begin{align*}
&\alpha _{k} = 4\left( { - 1} \right)^{k - 1}\left( {k - \ell - 1}
\right), \cr &\beta _{k} = \frac{1}{2}( - 1 )^{k - 1}( 2( \ell - ( k
- 4 ) )^{3} - 15( \ell - ( k - 4 ) )^{2} \cr &+  37( \ell - k + 4 )
- 30 ),  \cr &\gamma _{k} = \left( { - 1} \right)^{k}\left( {\ell -
k + 3} \right)\left( {\ell - k + 2} \right)\left( {\ell - k + 1}
\right), \cr &\delta _{k} = 2k\left( { - 1} \right)^{k}\left(
{4\ell^{2} + 2\ell\left( {2 - 3k} \right) + 2k^{2} - 3k + 1}
\right), \cr &S_{ - k} = 0, \: \: \: S_{0} = 1
\end{align*}
and $S_{k}$ is the symmetric function
\begin{equation}
S_{k} = \sum\limits_{r_{1} < r_{2} < ... < r_{k}} ^{\ell} {\wp
\left( {t_{r_{1} }}  \right)...\wp \left( {t_{r_{k}} } \right)}.
\label{20b}
\end{equation}

\noindent Using the first equation in the system  (\ref{19b}), we
can express $E$ through the quantities $\wp(t_{j})$ and derive the
following expression
\begin{equation}
E(t_{1},...,t_{\ell}) = \left( {2\ell - 1} \right)\sum\limits_{r =
1}^{\ell} {\wp \left( {t_{r}}  \right)}. \label{21b}
\end{equation}

Below we describe in detail the spectrum of the Schr\"{o}dinger
equation with the one-gap and two-gap Lame potentials.

\subsection{The one-gap Lame potential ($\ell=1$)}

The band edges are the following
\begin{equation}
\varepsilon_{1}=e_{3}, \: \: \varepsilon_{2}=e_{2},
 \: \: \varepsilon_{3}=e_{1} . \label{22b}
\end{equation}
The expression $\Lambda(x,E)$ for the one-gap potential is the
following
\begin{equation}
\Lambda(x,E)=\wp(i x +\omega )-E=\wp(i x +\omega )-\wp(t).
\label{23b}
\end{equation}
The spectrum of the Schr\"{o}dinger equation (\ref{3b}) with the
one-gap Lame potential in parametric form is the following
\begin{equation}
\left \{ \begin{array}[c]{rcl} k&=&\zeta(t)-\frac{\eta'}{\omega'}\:t, \\

E&=&\wp(t).
\end{array} \right. \label{24b}
\end{equation}
The expression for dependence $k$ on $E$ can be written in the form
of elliptic integral
\begin{equation}
k =k_{0} + \int\limits_{E_{0}} ^{E} {\frac{{z + \left( {{\eta
}'/{\omega} '} \right)}}{{2\sqrt {\left( {z - \varepsilon _{1}}
\right)\left( {z - \varepsilon _{2}} \right)\left( {z - \varepsilon
_{3}} \right)}} }dz}. \label{25b}
\end{equation}

\begin{figure}[!h]
\includegraphics[scale=0.8]{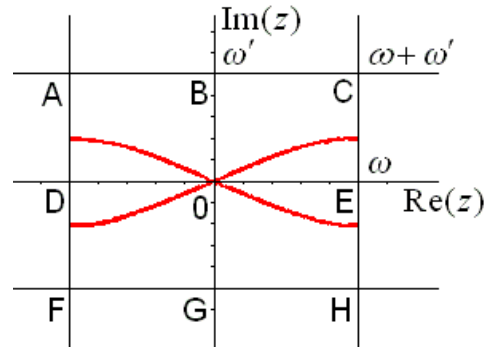}
\caption{\label{figure1} The Weierstrass function fundamental domain
on complex plane.}
\end{figure}

In  Fig.~\ref{figure1},  the complex plane and the fundamental
domain of the Weierstrass function $ACHF$ are represented. When  $E$
takes on values in the first band the parameter $t$ takes on values
in the line $AC$. The point B  corresponds to the minimum energy
value, the points $A$ and $C$ correspond to the maximum energy
value. When $E$ takes on values in the second band  the parameter
$t$ takes on values in the line $DE$. The points $D$ and $E$
correspond to the minimum energy value, the point $0$ corresponds to
the maximum energy value ($E=\infty$).

Let us examine limit cases.

In the limit case of free electrons the gap in the spectrum
vanishes. Let us put $\varepsilon _{2} = \varepsilon _{3} = -
\left( {{\eta} '/{\omega} '} \right)$ then $\varepsilon
_{1}=2\left( {{\eta} '/{\omega} '} \right)$ and
\begin{equation*}
k = \int\limits_{\varepsilon _{1} }^{E} {\frac{dz}{{2\sqrt {\left(
{z - \varepsilon _{1}} \right)} }}} = \sqrt {\left( {E - \varepsilon
_{1}}  \right)},
\end{equation*}
or $E = \varepsilon _{1} + k^{2}$.

In the limit case of a very wide gap, we pass to another variable $z
= \left( {1/2} \right) \cdot \left( {\varepsilon _{1} + \varepsilon
_{2}} \right) + \left( {1/2} \right) \cdot \left( {\varepsilon _{2}
- \varepsilon _{1}} \right)\varphi $ in the integral (\ref{25b}).
Using expression $e_{1}+e_{2}+e_{3}=0$ and introducing the
designation $A=(2E-\varepsilon _{1}-\varepsilon _{2})/(\varepsilon
_{2}-\varepsilon _{1}), \: \:A\in[-1;1] $ we derive the following
\begin{equation*}
k =\int\limits_{-1 }^{A}  \frac{{\frac{1}{2}  \left( {\varepsilon
_{1} + \varepsilon _{2}}  \right) +  {{\eta} '/{\omega} '}
+\frac{1}{2}  \left( {\varepsilon _{2} - \varepsilon _{1}}
\right)\varphi}}{{2\sqrt {(1-\varphi^{2})(\frac{3}{2}\varepsilon
_{3} - \frac{1}{2} \left( {\varepsilon _{2} - \varepsilon _{1}}
\right)\varphi)}} } d\varphi .
\end{equation*}
Expanding the expression under the integral to the order of
$\varepsilon _{2} - \varepsilon _{1}$ and restricting ourselves to
the zero approximation,  we obtain the spectrum in the limit case of
tight binding electrons.
\begin{equation*}
E = \frac{1}{2} \left( {\varepsilon _{1} + \varepsilon _{2} }
\right) - \frac{1}{2} \left( {\varepsilon _{2} - \varepsilon _{1}}
\right) \cos\left( {ka} \right),
\end{equation*}
where
\begin{equation*}
a=\frac{2\sqrt{\frac{3}{2}\varepsilon _{3}}}{\frac{1}{2}(\varepsilon
_{1} + \varepsilon _{2})+\frac{\eta'}{\omega'}} \: .
\end{equation*}

It is possible to determine effective masses at band edges.
According to the expression (\ref{25b}), we have
\begin{align*}
k = & \int\limits_{\varepsilon _{1}} ^{E}   \left(
\frac{{\varepsilon _{1} + \left( {{\eta }'/{\omega} '}
\right)}}{{\sqrt {\left( {\varepsilon _{2} - \varepsilon _{1}}
\right)\left( {\varepsilon _{3} - \varepsilon _{1}} \right)}} }
\cdot
\frac{1}{2\sqrt{z-\varepsilon _{1}}} \right. \nonumber  \\
&+ \left. O\left( (z-\varepsilon _{1})^{1/2}  \right) \right) dz,
\end{align*}
that leads to following expansions,
\begin{equation}
E=\varepsilon_{1}+\frac{k^{2}}{m_{1}^{*}}+O(k^{3});  \; \; \;\;
\frac{1}{m^{*}_{1}}=\frac{2(\varepsilon _{2}-\varepsilon
_{1})(\varepsilon _{3}-\varepsilon _{1})}{(\varepsilon _{1}+ {\eta
}'/{\omega} ')^{2}},   \label{26b}
\end{equation}
\begin{equation}
E=\varepsilon_{2}+\frac{k^{2}}{m_{2}^{*}}+O(k^{3}); \;\; \;\;
\frac{1}{m^{*}_{2}}=-\frac{2(\varepsilon _{2}-\varepsilon
_{1})(\varepsilon _{3}-\varepsilon _{2})}{(\varepsilon _{2}+ {\eta
}'/{\omega} ')^{2}}. \label{27b}
\end{equation}
Here we denote by $m_{1}^{*}$ and $m_{2}^{*}$ respectively the
effective masses at the bottom edge and at the top edge of the
first band. The effective mass $m_{1}^{*}$ is positive and
$m_{2}^{*}$ is negative. One can see that the both effective
masses depend on all the band edges.

\subsection{The two-gap Lame potential ($\ell=2$)}

For the two-gap Lame potential the band edges are the following,
\begin{align}
\varepsilon_{1}=-\sqrt{3g_{2}}, \; \;  \varepsilon_{2} =-3 & e_{1},
\;\;
 \varepsilon_{3} =-3e_{2}, \nonumber \\
&\varepsilon_{4}=-3e_{3}, \;\; \varepsilon_{5}=\sqrt{3g_{2}}.
\label{28b}
\end{align}
The expression $\Lambda(x,E)$ in this case is presented in  such a
way,
\begin{align}
&\Lambda(x,E) =18\wp^{2}(i x +\omega )-  6 E \wp(i x +\omega ) +2
E^{2}-\frac{9}{2}g_{2} \nonumber  \\&= 18 [\wp(i x +\omega
)-\wp(t_{1} )] \: [\wp(i x +\omega )-\wp(t_{2})] \nonumber  \\
      &=2\left[\frac{3}{2}\left(\wp(i x +\omega )- \sqrt{g_{2}-3\wp^{2}(i
x +\omega )} \right) -E\right] \nonumber  \\
&\times \left[\frac{3}{2}\left(\wp(i x +\omega
)+\sqrt{g_{2}-3\wp^{2}(i x +\omega )} \right)-E \right] .
\label{29b}
\end{align}
The spectrum of the Schr\"{o}dinger equation (\ref{3b}) with the
two-gap Lame potential in parametric form is the following
\begin{equation}
\left \{ \begin{array}[c]{rcl} k&=&\left(
\zeta(t_{1})-\frac{\eta'}{\omega'} t_{1}\right) + \left(
\zeta(t_{2})-\frac{\eta'}{\omega'}
t_{2}\right) , \\

\wp(t_{1})&=&\frac{E}{6}+\frac{1}{2\sqrt{3}}\sqrt{3g_{2}-E^{2}} , \\

\wp(t_{2})&=&\frac{E}{6}-\frac{1}{2\sqrt{3}}\sqrt{3g_{2}-E^{2}} .

\end{array} \right.  \label{30b}
\end{equation}
The expression for dependence $k$ on $E$ can be written in the form
of hyperelliptic integral
\begin{equation}
k=k_{o}+\int\limits_{E_{0}}^{E}\frac{\frac{3\sqrt{2}\eta'}
{\omega'}z+\frac{1}{\sqrt{2}}\left(2z\:^{2}-3g_{2} \right)
}{\sqrt{\left(z^{2}-3g_{2} \right) \left( 8z\:^{3}-18g_{2}z +54g_{3}
\right)}}dz , \label{31b}
\end{equation}

When $E$ takes on values in the first band the parameters  $t_{1}$
and $t_{2}$ take on values in the line $AC$ (Fig.~\ref{figure1}).
When $E$ takes on values in the second band the parameter $t_{1}$
takes on values in the line $AC$ and the parameter $t_{2}$ takes on
values in the line $DE$. In both cases $\wp(t_{1})$ and $\wp(t_{2})$
are real. The situation changes when $E$ takes on values in the
third band. In this case $\wp(t_{1})$ and $\wp(t_{2})$ are complex
conjugate ($k$ and $E$ remain real), $t_{1}$ and $t_{2}$ take on
values in the curves which are depicted inside fundamental domain
$ACHF$ (while constructing, we used $\omega =1, \: \omega'=i$).

With the help of the Hermite reduction, we can express the integral
(\ref{31b}) through the standard elliptic integrals of the first and
the second kinds  \cite{BE1, BE2} and represent the spectrum in
other parametric form. Let us put
\begin{equation*}
y=\frac{2z\:^{3}-b}{3(z\:^{2}-a)}, \; \: a=3g_{2}, \; \:  b=-54g_{3}
.
\end{equation*}
Using equalities
\begin{equation*}
\int\limits_{E_{0}}^{E}\frac{z d z}{\sqrt{\left(z^{2}-a \right)
\left( 8z\:^{3}-6az -b
\right)}}=\frac{1}{2\sqrt{3}}\int\limits_{Y_{0}}^{Y}\frac{d
y}{\sqrt{y^{3}-3ay+b}},
\end{equation*}
\begin{align*}
& \left. \int\limits_{E_{0}}^{E}\frac{(2z^{2}-a) d
z}{\sqrt{\left(z^{2}-a \right) \left( 8z\:^{3}-6az -b \right)}}
-\left[
\frac{1}{3}\sqrt{\frac{8z^{3}-6az-b}{z^2-a}}\right]\right|^{E}_{E_{0}} \\
& = \frac{1}{2\sqrt{3}}\int\limits_{Y_{0}}^{Y}\frac{y d
y}{\sqrt{y^{3}-3ay+b}},
\end{align*}
and putting $y=6 \wp(h)$, we can write the integral (\ref{31b}) in
terms of elliptic functions
\begin{equation}
\left \{ \begin{array}[c]{rcl} \wp
(h)&=&\frac{E^{3}+27g_{3}}{9E^{2}-27g_{2}} \; ,
\\

k&=& \zeta(h)-\frac{\eta'}{\omega'} h -
\frac{2}{3}\sqrt{\frac{(E+3e_1)(E+3e_2)(E+3e_3)}{E^2-3g_{2}}} \; .
\end{array} \right. \label{32b}
\end{equation}
In this form the spectrum of the two-gap potential is provided in
the paper. \cite{P}

By the same way as in the case of one-gap potential we derive
expressions for effective masses in the first band,
\begin{align}
&E=\varepsilon_{1} + \frac{k^{2}}{m_{1}^{*}}+O(k^{3}), \nonumber
\\
&\frac{1}{m^{*}_{1}} = \frac{2(\varepsilon_{2}-\varepsilon_{1})
(\varepsilon_{3}-\varepsilon_{1}) (\varepsilon_{4}-\varepsilon_{1})
(\varepsilon_{5}-\varepsilon_{1})}{
\left((\varepsilon_{1})^2+3({\eta }'/{\omega} ') \varepsilon_{1}
-\frac{3}{2}g_{2}  \right)^{2} }, \label{33b} \\
&E=\varepsilon_{2}+\frac{k^{2}}{m_{2}^{*}}+O(k^{3}), \nonumber \\
&\frac{1}{m^{*}_{2}} = -\frac{2(\varepsilon_{2}-\varepsilon_{1})
(\varepsilon_{3}-\varepsilon_{2}) (\varepsilon_{4}-\varepsilon_{2})
(\varepsilon_{5}-\varepsilon_{2})}{
\left((\varepsilon_{2})^2+3({\eta }'/{\omega} ') \varepsilon_{2}
-\frac{3}{2}g_{2}  \right)^{2} },
 \label{34b}
\end{align}
where $m_{1}^{*}$ and $m_{2}^{*}$ are effective masses at the bottom
and at the top band edges respectively.  One can see that   the
effective masses depend on all the band edges,  the same as in an
one-gap case. In Table~\ref{table1}, the effective mass values are
provided for different values of the parameter $ |\tau| $, we put a
period of the potential $T=|2\omega'|=2$.

\begin{table} [!h]
\caption{\label{table1}Inverse effective masses  for the first band
in case of the two-gap Lame potential.}
\begin{ruledtabular}
\begin{tabular}{ccc}
$\tau$ & $1/m^{*}_{1}$ & $1/m^{*}_{2}$ \\
\hline 0.55 & 1.99 & -47.34    \\  0.74 & 1.89 & -10.03   \\
 1 & 1.25 & -2.13    \\   1.23 & 0.57 & -0.67
\\  1.52 & 0.15 & -0.16  \\     2.18 & 0.005 & -0.005
\\
\end{tabular}
\end{ruledtabular}
\end{table}

In Table~\ref{table1}, one can see that  with the increase of the
value of $|\tau|$, the values of $|1/m^{*}_{1}|$ and $|1/m^{*}_{2}|$
decrease. When the band degenerates into an energy level that
corresponds to the limit case of tight binding electrons, $m^{*}_{1}
\rightarrow \infty,\; m^{*}_{2} \rightarrow -\infty$. We obtain 
so called "heavy" electrons. In the other limit case of free
electrons, $m^{*}_{1}$ tends to the mass of free electron, i.e.
$m^{*}_{1}\rightarrow 1/2$. And since the gap width vanishes,
$m^{*}_{2} \rightarrow - 0 $ in this case.

In Fig~\ref{figure2}, we provide the graphs $E=E(k)$ and $U(x)$ for
the two-gap potential corresponding to different  values of the
parameter $ \left| \tau \right|=\left|{\omega'}\:/{\omega}\right|$.
As $|\tau|$ increases, the width of the gaps increase. There are the
limit cases of very narrow gaps and very wide gaps in this figure.
With increase of $|\tau|$, the second band also degenerates into an
energy level.

\begin{figure}[!h]
\includegraphics[scale=0.5]{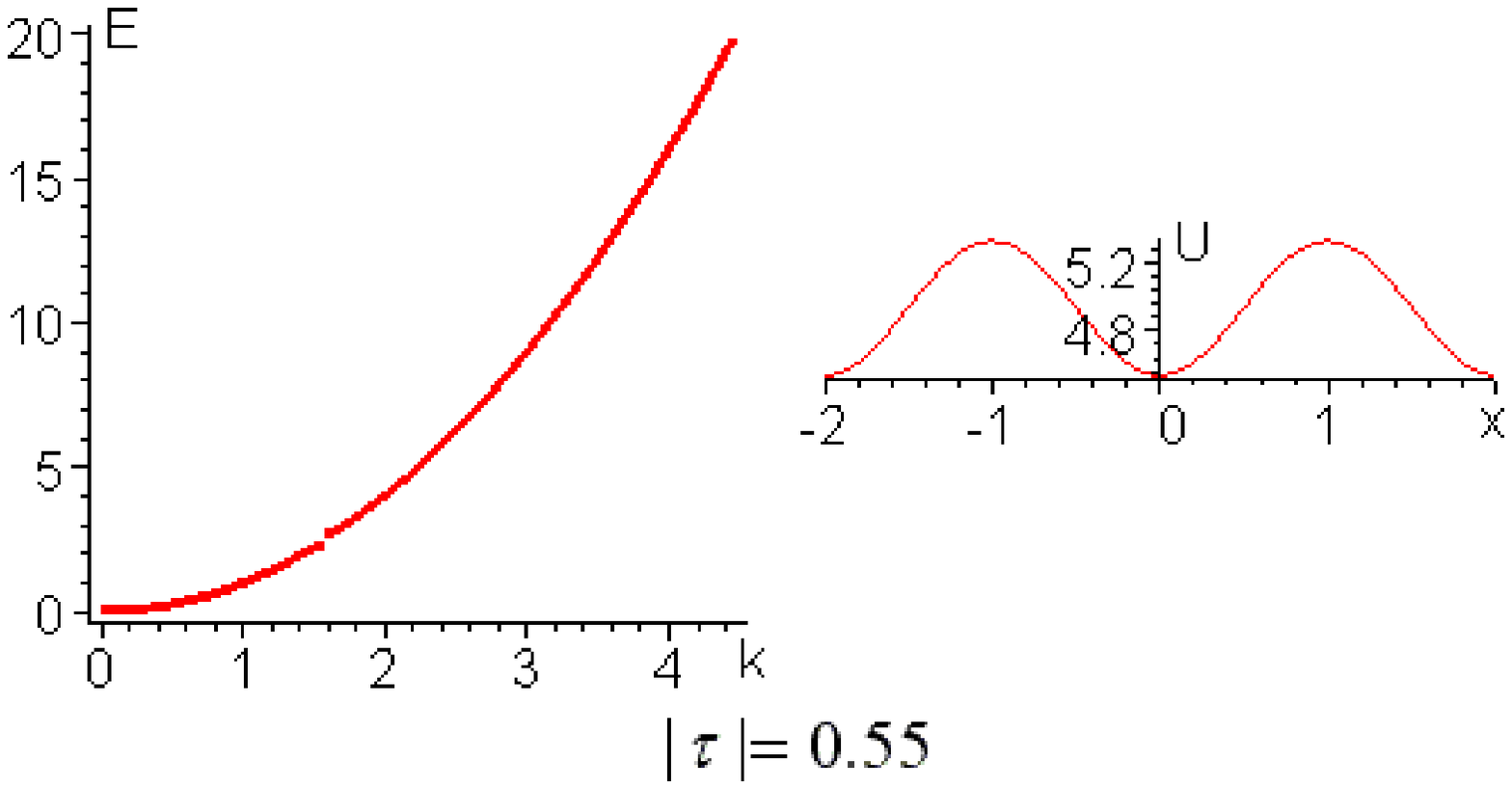}
\end{figure}
\begin{figure}[!h]
\includegraphics[scale=0.5]{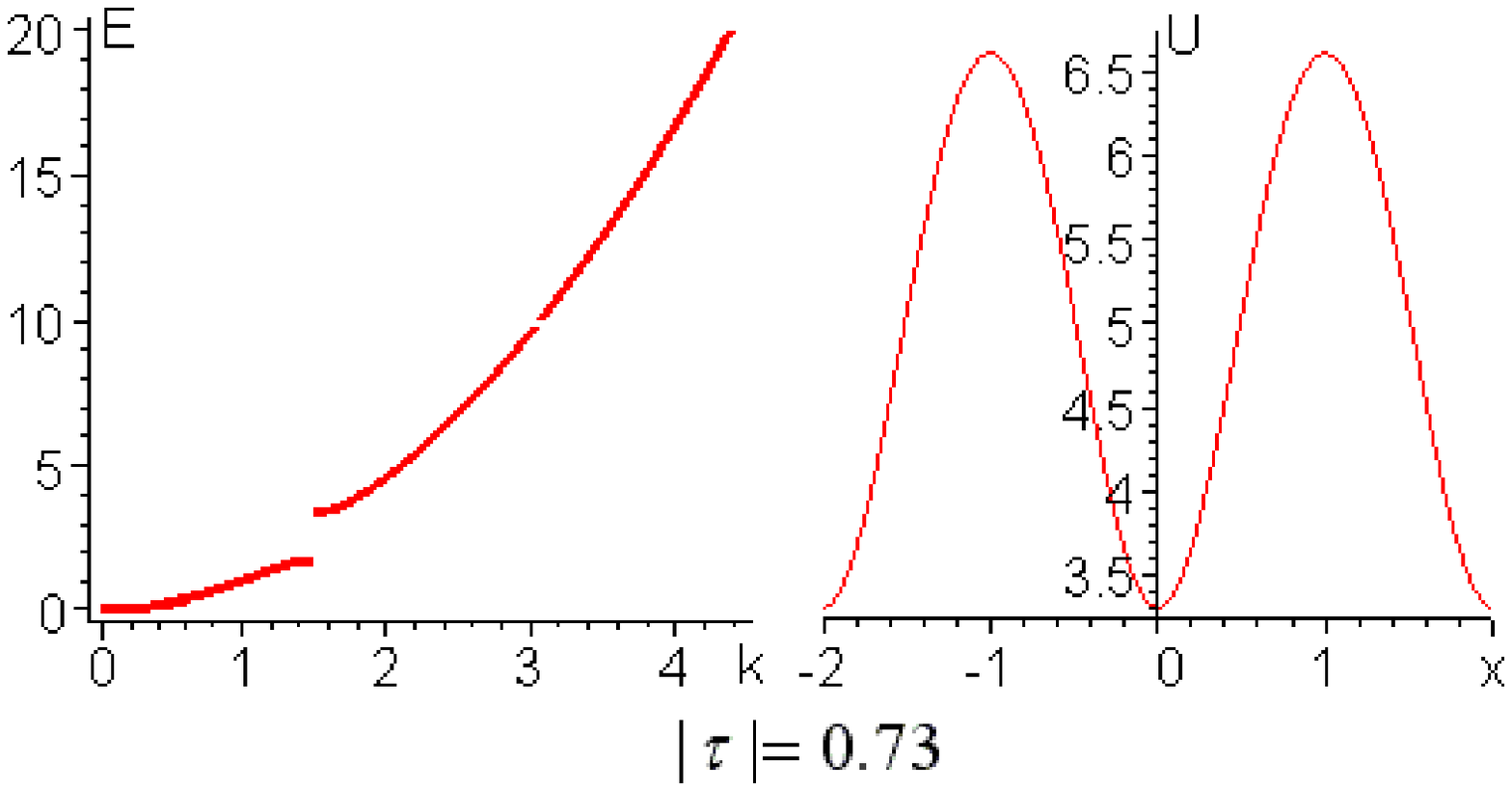}
\end{figure}
\begin{figure}[!h]
\includegraphics[scale=0.5]{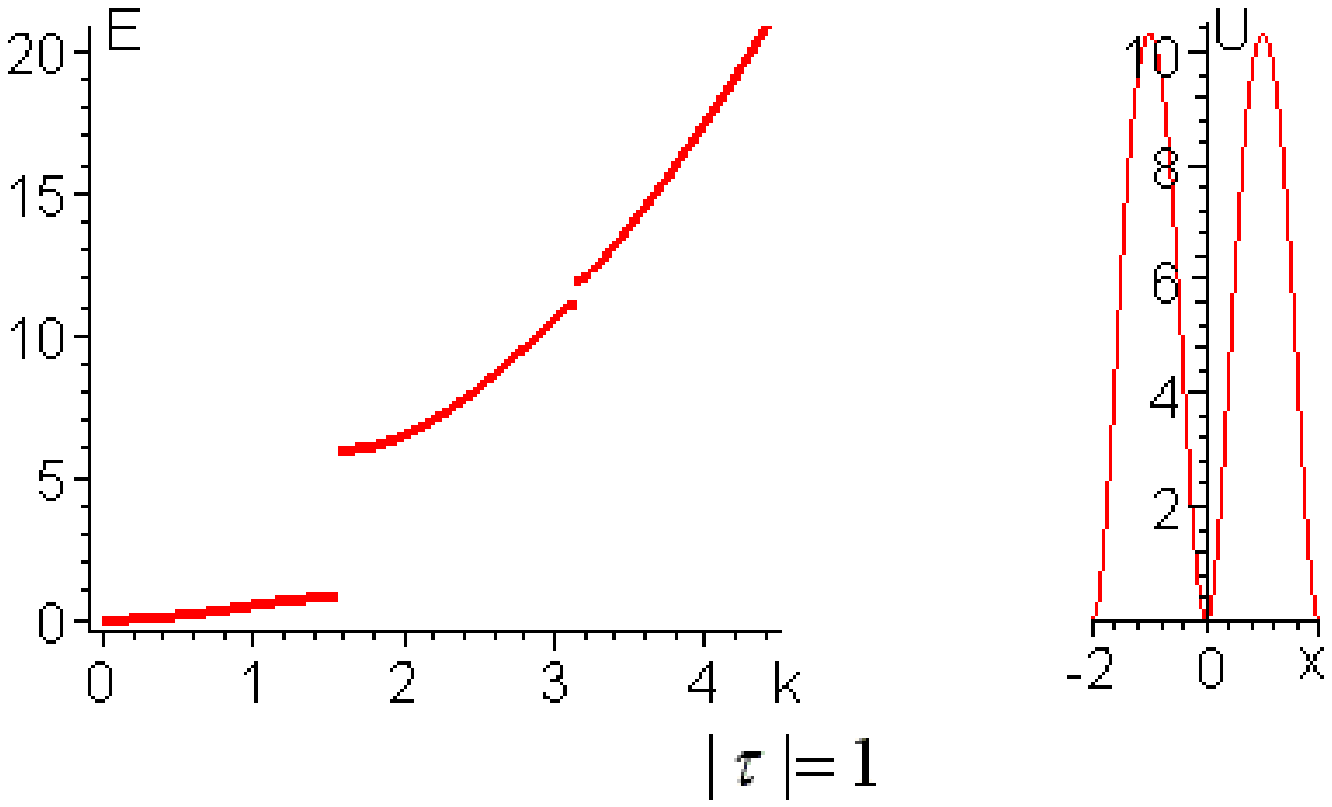}
\end{figure}
\begin{figure}[!h]
\includegraphics[scale=0.5]{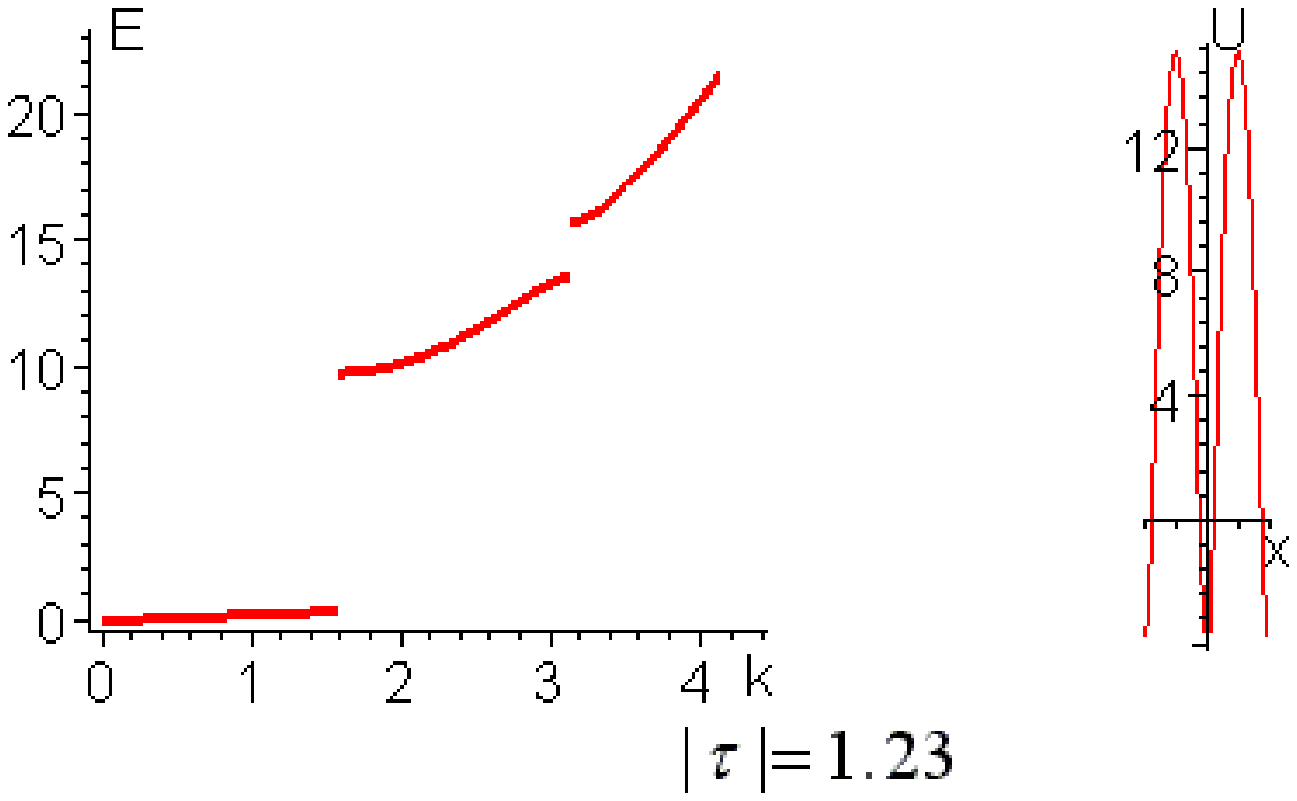}
\end{figure}
\begin{figure}[!h]
\includegraphics[scale=0.5]{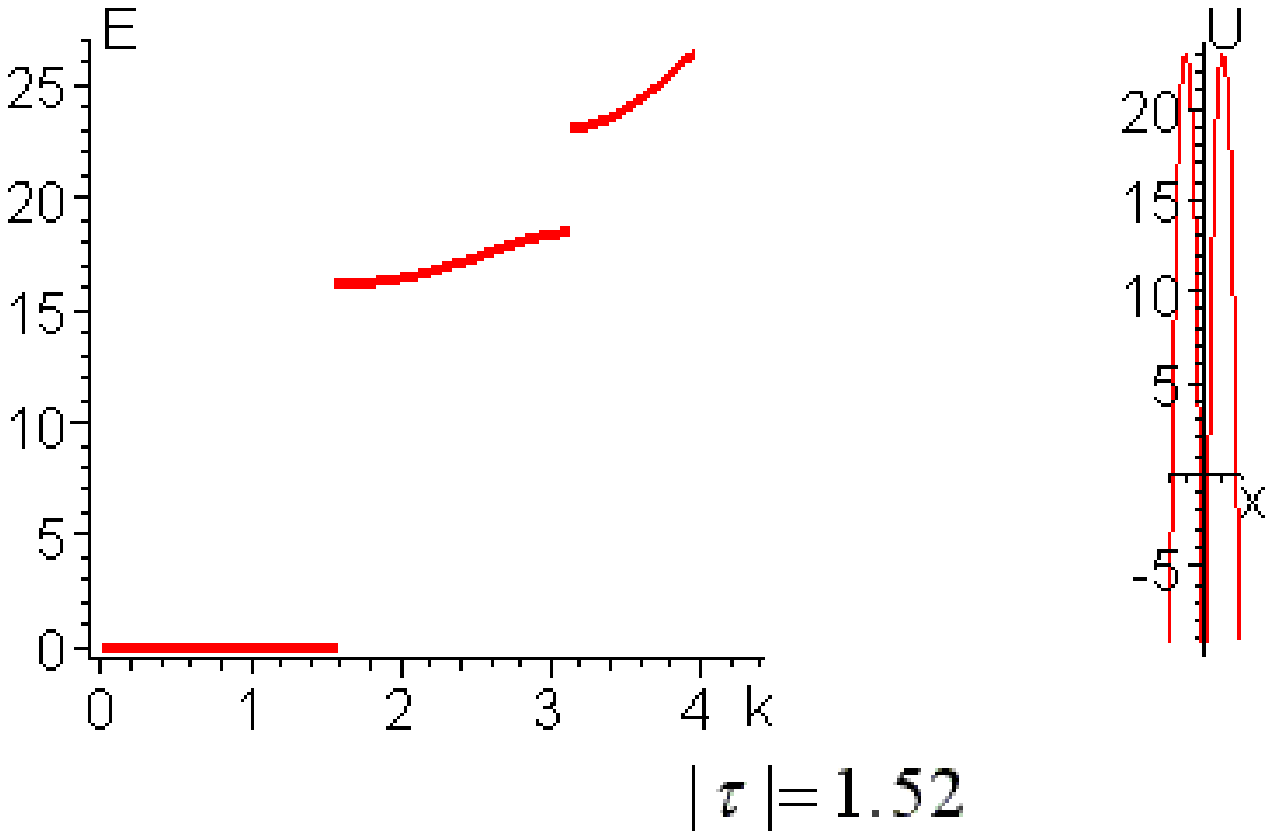}
\caption{\label{figure2}  Dispersion relation $E=E(k)$ and potential
$U(x)$ in case of the two-gap Lame potential for different values of
the parameter $ \left| \tau
\right|=\left|{\omega'}\:/{\omega}\right|$. }
\end{figure}

\section{The separable finite-gap Lame potential}
The one-electron separable Hamiltonian has the form
\begin{equation}
H=\sum\limits_{j=1}^{n} H_{j}=\sum\limits_{j=1}^{n} \left(
-\frac{\partial^{2}}{\partial x_{j}^{2}} +U_{j}(x_{j}) \right) .
\label{35c}
\end{equation}
In general case $n\in \mathbb{N},$ but we consider below only the
three dimensional case, $n=3.$ We will use the finite-gap Lame
potentials for the potentials $U_{j}(x_{j})$. In general case of the
orthorhombic lattice every one-dimensional potential $U_{j}(x_{j})$
is characterized by its own parameters $\omega, \: \omega'$,
\begin{align}
&H=-\partial_{x}^{2}-\partial_{y}^{2}-\partial_{z}^{2}-\ell_{x}(\ell_{x}+1)\wp(i
x+\omega_{x})\nonumber \\ &-\ell_{y}(\ell_{y}+1)\wp(i
y+\omega_{y})-\ell_{z}(\ell_{z}+1)\wp(i z+\omega_{z}) . \label{36c}
\end{align}

\begin{figure}[!h]
\includegraphics[scale=0.65]{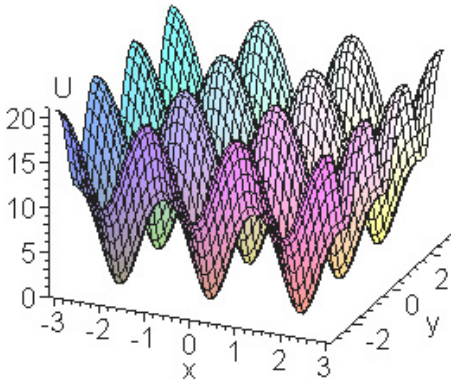}
\caption{\label{figure3} The three dimensional one-gap Lame
potential for fixed value of the variable $z$, we put $z=0$,
$\omega_{x}=\omega_{y}=1, \; \omega'_{x}=\omega'_{y}=i$. }
\end{figure}

In Fig.~\ref{figure3}, the three dimensional one-gap potential for
$z=0$ is represented with the following half-periods:
$\omega_{x}=\omega_{y}=1, \; \omega'_{x}=\omega'_{y}=i$.

An eigenfunction  of the Hamiltonian  (\ref{36c}) is a product of
one-dimensional eigenfunctions of every $H_{j}$,
\begin{equation}
\Psi_{\mathbf{I},\; \mathbf{k}}(\mathbf{x})=
\prod\limits_{j=1}^{3} \Psi_{I_{j}, \; k_{j}}(x_{j}), \label{37b}
\end{equation}
where $\textbf{I} = (I_{1}, I_{2}, I_{3}), \;$   $I_{j} \in \{
1,..,\ell_{j}\}$ is a band number. The expression for energy
consists of three terms,
\begin{equation}
E_{\mathbf{I}}(\mathbf{k})=E_{I_{1}}(k_{1})+E_{I_{2}}(k_{2})+E_{I_{3}}(k_{3})
.  \label{38b}
\end{equation}
The Fermi surface is described by the following equation
\begin{equation*}
E_{F}=E_{\mathbf{I}}(\mathbf{k}+\mathbf{b}),
\end{equation*}
were $E_{F}$ is the Fermi energy. Here the wavevector $\mathbf{k}$
takes on values in the Brillouin zone, every solution of this
equation for a prescribed value of dual lattice vector $\mathbf{b}$
defines one of sheets of the Fermi surface. The set of dual lattice
vectors corresponds to their set in Fourier series expansion of a
separable potential. If the real electron potential differs a bit
from the separable one, then the corresponding corrections can be
taken into account using the perturbation theory.

Let us examine the limit  case of  very  narrow gaps. Since the
potential periods $T_{j}=2|\omega_{j}' |$ are fixed, let $\omega_{j}
\rightarrow \infty $. Then, potentials $U_{j}(x)$ tend to constants,
the dependence of energy on wavevector becomes quadratic and the
Fermi surface sheets are similar to pieces of spherical surface.
Thus, our model is the considerable generalization of the Harrison
method that was successfully used to construct the Fermi surfaces of
metals. \cite{H}

Let us examine the other limit  case of   very wide gaps. Since we
fixed the potential periods $T_{j}=2|\omega_{j}'|$, let $\omega_{j}
\rightarrow 0$. Then, $e_{2}^{(j)}-e_{3}^{(j)} \rightarrow 0$. It is
obvious that in this case all the bands degenerate into levels. The
finite-gap spectrum of tight binding electrons is similar to the
spectrum that is obtained in the well-known LCAO method. In this
case function $\Psi _{I_{0}, \; 0} \left( {x} \right); \: {I_{0} =
\left( {0,0,0} \right)}$ corresponds to $s$-state. The three
functions $\Psi _{I_{1}, \; 0} \left( {x} \right), $ $ \Psi _{I_{2},
\; 0} \left( {x} \right), $ $ \Psi _{I_{3}, \; 0} \left( {x}
\right);$ $ I_{1} = ( {1,0,0} ),$ $ I_{2} = ( {0,1,0} ),$ $ I_{3} =
\left( {0,0,1} \right)$ correspond to $p$-state as they are
proportional to $x$, $y$, $z$ respectively when  the space variables
tend to zero. The six functions $\Psi _{I_{4}, \; 0} ( {x} )$, $\Psi
_{I_{5}, \; 0} ( {x} )$, $\Psi _{I_{6}, \; 0} ( {x})$, $\Psi
_{I_{7}, \; 0} ( {x})$, $\Psi _{I_{8}, \; 0} ( {x} )$, $\Psi
_{I_{9}, \; 0} ( {x} )\:$; $I_{4} = \left( {1,1,0} \right)$, $I_{5}
= \left( {1,0,1} \right)$, $I_{6} = \left( {0,1,1} \right)$, $I_{7}
= \left( {2,0,0} \right)$, $I_{8} = \left( {0,2,0} \right)$, $I_{9}
= \left( {0,0,2} \right)$ split into five functions that correspond
to  $d$-state and one function corresponds to the $s$-state since
these functions are proportional to $xy,\; xz, \; yz$, $x^{2} - a,
\; y^{2} - b, \; z^{2} - c$ respectively when the space variables
tend to zero. \cite{BBK2}

\begin{figure}[!h]
\includegraphics[scale=0.5]{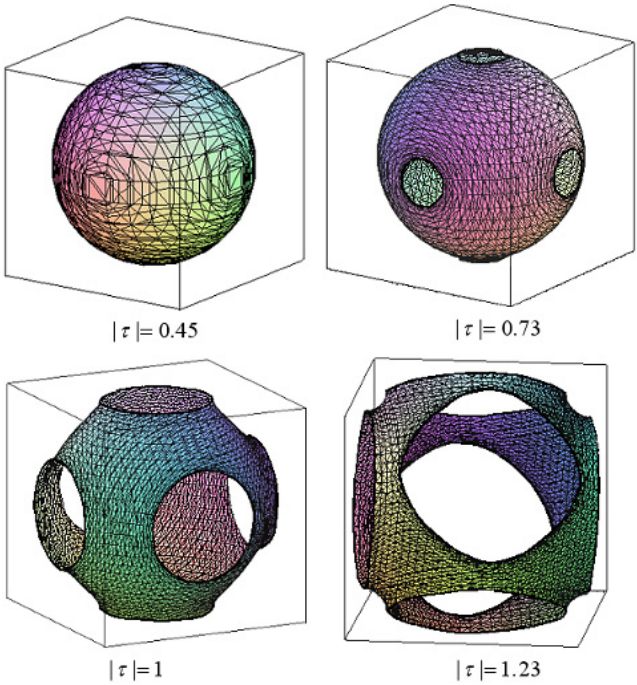}
\caption{\label{figure4} The Fermi surface in the
first band for different values of the parameter  $ \left| \tau
\right|=\left|{\omega'}\:/{\omega}\right|$, the Fermi energy is
$E_{F}=2.43$.}
\end{figure}

\begin{figure}[!h]
\includegraphics[scale=0.5]{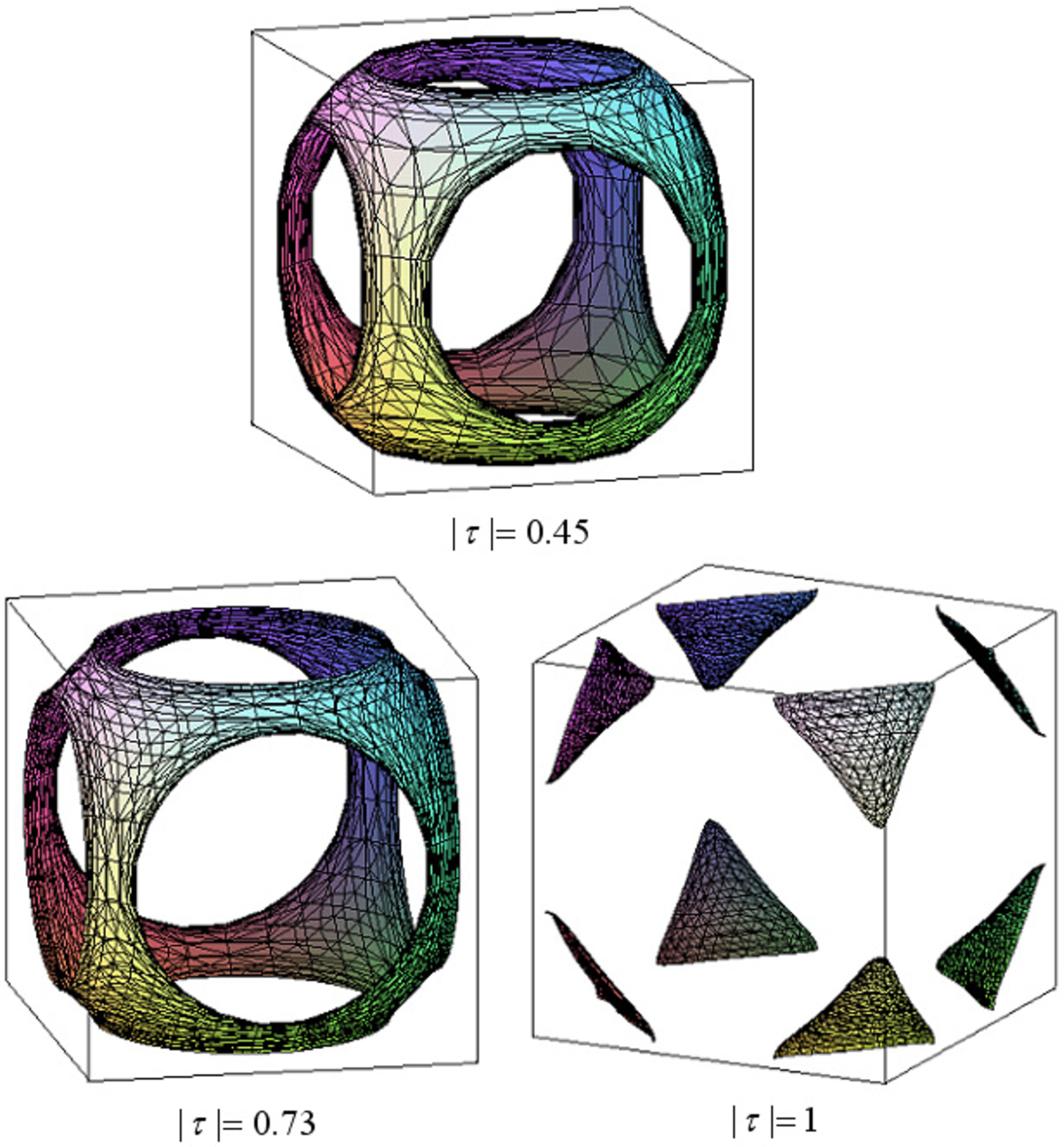}
\caption{\label{figure5} The Fermi surface in the
first band for different values of the parameter  $ \left| \tau
\right|=\left|{\omega'}\:/{\omega}\right|$, the Fermi energy is
$E_{F}=3.8$.}
\end{figure}

The Fermi surfaces for cubic lattice are represented in
Fig.~\ref{figure3}-\ref{figure5}. They are constructed using the
separable one-gap Lame potential for different values of the
parameter $|\tau|$. In Fig.~\ref{figure3}, the Fermi surfaces in the
first band are represented at $E_{F}=2.43$. The Fermi surface sheets
don't appear in the second band at this value of $E_{F}$. In
Fig.~\ref{figure4}, the open Fermi surfaces are represented at
$E_{F}=3.8$ in the first band. In Fig.~\ref{figure5}, the closed
Fermi surfaces are represented at the same value of $E_{F}$ in the
second band.

\begin{figure}[!h]
\includegraphics[scale=0.5]{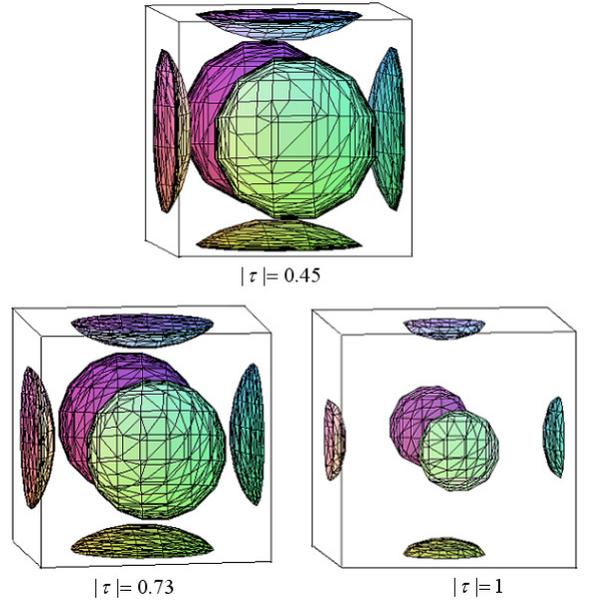}
\caption{\label{figure6} The Fermi surface in the
second band for different values of the parameter  $ \left| \tau
\right|=\left|{\omega'}\:/{\omega}\right|$, the Fermi energy is
$E_{F}=3.8$.}
\end{figure}

\section{The Lifshits singularities of the  electron thermodynamic potential
and the one-gap Lame potential}

It is known that the density of states in a crystal
\begin{equation*}
\nu(E)=\frac{dN(E)}{dE}=\frac{2V}{(2\pi)^3} \int
\limits_{E(\mathbf{k})=E}\frac{dS}{| \mathbf{\nabla} E( \mathbf{k})
|} ,
\end{equation*}
as a function of the variable $E$ has singularities which are named
the Van Hove ones.\cite{V} They occur when the electron group
velocity vanishes, $\mathbf{\nabla} E( \mathbf{k_{c}})=0$. Since $E
(\mathbf{k})$ is a smooth function of the  wavevector $\mathbf{k}$,
it can be expanded into the Taylor series about a critical
wavevector $\mathbf{k_{c}}.$ Restricting ourselves to quadratic
terms, we obtain
\begin{equation*}
E=E_{c}+\alpha_{1}\xi_{1}^{2}+\alpha_{2}\xi_{2}^{2}+\alpha_{3}\xi_{3}^{2}+O(\xi^{3}),
\end{equation*}
where $\mathbf{\xi}=\mathbf{k}-\mathbf{k_{c}}$. Depending on signs
of the coefficients ${\alpha_{1}, \alpha_{2}, \alpha_{3} }$ the
four types of critical points are distinguished, they are a
minimum value, a maximum value and two types of saddle points. In
a three dimensional crystal the function $\nu(E)$ at the critical
energy $E_{c}$ is continuous but its derivative at this point has
infinite discontinuity.

The function $E(\mathbf{k})$ is periodic with respect to every
component of the wavevector $\mathbf{k}=(k_{x},k_{y},k_{z}),$ it
means the energy $E(k_{x},k_{y},k_{z})$ is defined on three
dimensional torus. This function  has a denumerable set of branches
and every one of them corresponds to a certain energy band. The
function $E(k_{x},k_{y},k_{z})$ is differentiable and regular so,
according to elementary facts of topology, for each of the branches
the number of critical points relates to the Betti numbers of the
manifold, where the function $E$ is being defined. Thus, every
branch of the function $E(k_{x},k_{y},k_{z})$ has one maximum value,
one minimum value and three saddle points of every type. But in a
real crystal the bands may overlap and as a result the singularities
of the function $\nu(E)$, that are related to different branches,
may compensate each other.

In 1952 Van Hove  examined singularities of the elastic frequency
distribution of a crystal (that are similar to  singularities of the
electron density of states) and showed that for a three dimensional
crystal the frequency distribution function has at least one saddle
point of every types and its derivative takes on the value $-\infty$
at the upper end of the spectrum. \cite{V}

In 1960 I.M. Lifshits  connected singularities of the  density of
electron states  with a variation of the Fermi surface topology.
\cite{L} He showed that at zero temperature the electron phase
transitions occur at the special energies, they were later named the
Lifshits electron phase transitions of $2 \frac{1}{2}$ kind.
Singularities of the conduction electron spectrum at the point of
such a transition lead to anomalies of thermodynamic and kinetic
quantities. When temperature increases, this  transition smoothes
out.

At present, the Van Hove singularities are considered as one of the
reasons for the origin of high-temperature superconductivity.
\cite{M}

\begin{figure}
\includegraphics[scale=0.65]{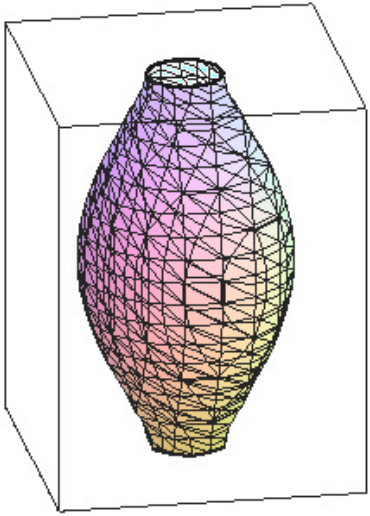}
\caption{\label{figure7} The Fermi surface in the
first band in the form of a goffered cylinder, we put
$\tilde{\omega}=1, \; \tilde{\omega} ' =0.65 i; \; \omega=0.3 , \;
\omega'=0.5 i$.}
\end{figure}

Let us consider the Schr\"{o}dinger equation with the one-gap
separable Lame potential that corresponds to the tetragonal
Bravais lattice,
\begin{align*}
-\mathbf{\Delta} &\Psi+  W(x,y,z)\Psi = E \Psi, \\  W(x, y, z)= & -
2
\wp(i x + \tilde{\omega} | \tilde{\omega}, \tilde{\omega}' \: ) \\
- 2 \wp(i y + \tilde{\omega} & | \tilde{\omega}, \tilde{\omega}' \:
)- 2 \wp(i z + \omega | \omega, \omega' \: ) - e_{3}+ 4
\tilde{e}_{1} .
\end{align*}
In order to describe the Lifshits electron transformation,  it is
enough to examine the case when the Fermi surface is open  only at
two opposite sides of the Brillouin zone. Let us choose half periods
and the Fermi energy so that the Fermi surface has the form of a
goffered cylinder which is prolate along the direction $k_{z}$, as
in Fig.~\ref{figure7}. In this case we have to put $\tilde{\omega}
> \omega, \: | \tilde{\omega}' \:| >  | \omega' \: |
$ (while constructing, we used $\tilde{\omega}=1, \; \tilde{\omega}
' =0.65 i; \; \omega=0.3 , \; \omega'=0.5 i$). Then, in order to
simplify computations, we can expand potential with respect to the
variables $x, \: y$ and restrict ourselves to zero approximation.

The spectrum of the Schr\"{o}dinger operator with the potential
\begin{equation}
U(x,y,z)=- e_{3} -2\wp(i z + \omega | \omega, \omega' \: )
\label{39d}
\end{equation}
looks as follows:
\begin{equation} \left \{ \begin{array}[c]{rcl}
E(k_{x},k_{y},h_{z})  &=& k_{x}^2+k_{y}^2 -e_{3}+\wp\left(h_{z}
\right)  ,
\\

k_{z}(h_{z}) &=&  \zeta(h_{z}) -\frac{\eta'}{\omega'} \; h_{z} \; .
\end{array} \right. \label{40d}
\end{equation}
The function $E(k_{x},k_{y},k_{z})$ has two critical points in the
first band that are a minimum value and a saddle point and one
critical point in the second band that is a minimum value. We will
consider below only the first band.

One can see  that $E|_{\mathbf{k}=0}=0$. Let us put
$K^2=k_{x}^2+k_{y}^2$. For the first band $h_{z}=t+\omega', \: t \in
[-\omega; \: \omega]$. Since, the function $E(\mathbf{k})$ is even,
it is enough to consider the case when $t \in [0; \: \omega]$. The
expression for the number of states is the following
\begin{align}
&N(E)=\frac{2V}{(2\pi)^3} \int dk_{x} dk_{y} dk_{z}=
\frac{2V}{(2\pi)^3}\int{\pi K^2 d k_{z}} \nonumber \\    &=
\frac{2V}{(2\pi)^3}\int{
\pi \left(E+e_{3} - \wp\left(h_{z} \right) \right) d k_{z}}  \nonumber  \\
&= \frac{2V}{(2\pi)^3}\int{ \pi \left(E+e_{3} - \wp\left(h_{z}
\right) \right) \left( - \wp\left(h_{z} \right)
-\frac{\eta'}{\omega'} \right) d h_{z}}, \label{41d}
\end{align}
where $V$ is dimensionless volume.

It easy to show that at the energy $E_{c}=-e_{3}+e_{2}$ the Fermi
surface opens on the boundary of the Brillouin zone which is
orthogonal to the axis $k_{z}.$ In this case we can express the
number of states (\ref{41d}) in such a form,
\begin{widetext}
\begin{align}
&N(E)=\frac{V}{2\pi^2}\int\limits_{0}^{t_{0}} { \left(E+e_{3} -
\wp\left(t+\omega' \: \right) \right) \left( - \wp\left(t+\omega' \:
\right) -\frac{\eta'}{\omega'} \right) d t}\cr &=\frac{V}{2\pi^2}
\left[ \left( \frac{g_{2}}{12}-\frac{\eta'}{\omega'}
(E+e_{3})\right)t_{0} +  \left( \zeta (t_{0}+\omega' \:)-\eta' \:
\right) \left( (E+e_{3})- \frac{\eta'}{\omega'} \right)
+\frac{1}{6}\wp'\left(t_{0}+\omega'  \: \right)
 \right]. \label{e1}
\end{align}
\end{widetext}

Here we have introduced the parameter $t_{0}$ as follows.

\textbf{A.} Under condition  $ 0 \leq  E < E_{c}$ we define the
parameter $t_{0}$ by the equality
\begin{equation*}
E+e_{3}=\wp\left(t_{0}+\omega' \right) .
\end{equation*}
According to this equality, the function $t_{0}=t_{0}(E)$ is
differentiable and has a derivative
\begin{equation*} \frac{d
t_{0}}{dE}=\frac{1}{\wp'\left(t_{0}+\omega' \right)}.
\end{equation*}

\textbf{B.}  Under condition  $E > E_{c}$ we define the parameter
$t_{0}$ by the equality $t_{0}=\omega$.

Differentiating this expression for the number of states in $E$
and using the definition of parameter $t_{0}$ we obtain the
electron density of states,
\begin{align}
\nu(E)&=\frac{d N(E)}{d E}=\frac{V}{2\pi^2}\int\limits_{0}^{t_{0}} {
\left( - \wp\left(t+\omega' \: \right) -\frac{\eta'}{\omega'}
\right) d t}\cr & =\frac{V}{2\pi^2}\left(  \zeta (t_{0}+\omega' \:)
-\frac{\eta'}{\omega'} \: (t_{0}+ \omega' \; ) \right) \cr &=\left
\{
\begin{array}[c]{rcl} \frac{V}{2\pi^2} | \mathbf{k}(E) |,& & E <
E_{c}
\\
\frac{V}{2\pi} \cdot \frac{1}{2| \omega' |},& &  E > E_{c}
\end{array} \right.. \label{43d}
\end{align}
Even though this expression was derived for the one-gap potential,
it is correct in a case of any  $\ell-$gap potential.

Let us expand the function $\nu(E)$ under the condition $E <
E_{c}$ in the $\sqrt{E}$ powers. We derive the following formula,
\begin{align}
\nu(E)=\frac{  V \;   E^{1/2}}{2 \pi^2 C^{1/2}}  +O (E^{3/2}),
\quad C=\frac{6
e_{3}^{2}-\frac{g_{2}}{2}}{2(e_{3}+\frac{\eta'}{\omega'})^{2}}.
\label{44d}
\end{align}
It is obvious that the first term in the expansion  (\ref{44d})
corresponds to the function
$E(k_{x},k_{y},k_{z})=k_{x}^{2}+k_{y}^{2}+C_{z} k_{z}^{2}$.

Let us calculate the first derivative of the density  of states.
We should distinguish two cases.

\textbf{A.}  Under condition $ 0\leq E < E_{c}$ we obtain the
expression
\begin{align}
&\frac{d \nu(E)}{d E}=-\frac{V}{2\pi^2} \cdot \frac{
\wp\left(t_{0}+\omega' \: \right) +\frac{\eta'}{\omega'}}{\wp '
\left(t_{0}+\omega' \: \right)} \nonumber \\  &= -\frac{V}{4 \pi^2}
\cdot
\frac{\frac{\eta'}{\omega'}+e_{3}+E}{E^{1/2}(e_{1}-e_{3}-E)^{1/2}}
\cdot \frac{1}{(E_{c}-E)^{1/2}} . \label{45d}
\end{align}

\textbf{B.}   Under condition $E > E_{c}$ we obtain the expression
\begin{align}
\frac{d \nu(E)}{d E}=0. \label{46d}
\end{align}
The singularity at the point $E=e_{1}-e_{3}$ corresponds to the
Fermi surface appearance in the second band. The singularity at
the point $E=0$  corresponds to the minimum value of $E$ in the
first band (the Fermi surface appears in the center of the
Brillouin zone). The singularity at the point $E=E_{c}$
corresponds to the saddle point (the Fermi surface opens on the
boundary of the Brillouin zone). In all three cases the
singularities have positive signs. We will consider further only
the case $E=E_{c}$, it is obvious that the other two cases can be
considered similarly.

Let us introduce the following notation:
\begin{align*}
&\frac{\alpha}{2} = -\frac{V}{4\pi^2} \cdot
\frac{\frac{\eta'}{\omega'}+e_{2}}
 {(e_{2}-e_{3})^{1/2}(e_{1}-e_{2})^{1/2}}
 \cr
&= \frac{V}{4\pi^2} \cdot \frac{| \frac{\eta'}{\omega'}+e_{2} |}
 {(e_{2}-e_{3})^{1/2}(e_{1}-e_{2})^{1/2}}, \cr
&A = \frac{V}{2 \pi^2} \cdot \left( \frac{g_{2}}{12}\omega+
\frac{\pi}{2|\omega' \: |}e_{3}-\frac{\eta \eta'}{\omega'} \:
\right), \cr &B=\frac{V}{2\pi} \cdot \frac{1}{2|\omega' \; |}, \cr
&C= -\frac{V}{2\pi^2}  \left[ \frac{e_{2}^{2}}{2} \cdot
\frac{\pi}{2|\omega'|}+
e_{2}\left(\frac{g_{2}}{12}\omega-\frac{\eta'}{\omega'}\eta \right)
+\frac{3}{40}g_{2}\eta\right. \cr
&\left.-\omega
\left(\frac{\eta'}{\omega'} \cdot \frac{g_{2}}{12}+\frac{1}{20}g_{3}
\right) \right].
\end{align*}
In order to compute  the thermodynamics potential
$\Omega(T,V,\mu)$ we expand the following expressions $d \nu(E)/d
E, \:$ $\nu(E), \:$ $N(E)$  in the $(E_{c}-E)^{1/2}$ powers. We
derive the following expressions:
\begin{equation}
\frac{d \nu(E)}{d E}=\left \{ \begin{array}{lr}
 \frac{\alpha}{2} \cdot \frac{1}{(E_{c}-E)^{1/2}} \\ \hspace{1.2cm} +O\left( (E_{c}-E)^{1/2}
 \right),&  E < E_{c},
\\
0,&   E > E_{c},
\end{array} \right. \label{47d}
\end{equation}
\begin{equation}
\nu(E)=\left \{ \begin{array}{lr}
 B - \alpha  (E_{c}-E)^{1/2} \\ \hspace{1.5cm}  +O\left( (E_{c}-E)^{3/2}
 \right),&  E < E_{c},
\\
B,&   E > E_{c},
\end{array} \right. \label{48d}
\end{equation}
\begin{equation}
N(E)=\left \{ \begin{array}{lr} A+ B E + \frac{2}{3} \alpha
(E_{c}-E)^{3/2} \\ \hspace{1.3cm} +O\left( (E_{c}-E)^{5/2}
 \right),&  E < E_{c},
\\

A+B E ,&   E > E_{c}.
\end{array} \right. \label{49d}
\end{equation}

The coefficient $\alpha$ is expressed in terms of the effective
mass $m_{2}^{*}$ along the $k_{z}$ direction at the top edge of
the first band,
\begin{align}
\alpha = \frac{V}{\sqrt{2}\pi^2} \sqrt{-m_{2}^{*}} \; . \label{d}
\end{align}
Hence the coefficient $\alpha$ depends on all the band edges and
$\alpha \in [0;\infty)$. When the gap width decreases on boundary of
the Brillouin zone, which is orthogonal to the $k_{z}$ axis, the
value of the  coefficient $\alpha$ decreases also. In the limit case
of extremely wide gap the parameter $\alpha \rightarrow\infty $. In
this case the problem from a three-dimensional one becomes
two-dimensional and we should use other speculations. In the other
limit case when the gap width vanishes the parameter $\alpha
\rightarrow 0 $. It is obvious: when the gap width equals zero the
value $E_{c}$ is not a critical point.

Thermodynamic potential at a normal metal is defined by the
following expression
\begin{align}
\Omega(T,V,\mu) &= - \int\limits_{0}^{\infty}\frac{N(\varepsilon)d
\varepsilon }{1+exp\left(\frac{\varepsilon-\mu}{T} \right)} ,   \cr
\Omega(0,V,\mu)&=\Omega_{0} = - \int\limits_{0}^{\mu}N(\varepsilon)d
\varepsilon \; . \label{e2}
\end{align}
Using the expression (\ref{e1}) we can calculate  the integral
(\ref{e2}) and find the analytical expression for $\Omega(0,V,\mu)$,
but this expression will be quite a complicated one and  we do not
adduce it.

Using the formula (\ref{49d}) we present the electron thermodynamic
potential $\Omega_{0}$ in a neighborhood of the critical value of
energy in such a form,
\begin{widetext}
\begin{equation}
\Omega_{0}=\left \{ \begin{array}[c]{lcl} C+A E_{c}+\frac{1}{2} B
E_{c}^2 -A \mu-\frac{1}{2} B \mu^2+ \frac{4}{15}\alpha
(E_{c}-\mu)^{5/2}+O \left( (E_{c}-\mu)^{7/2} \right)
 ,& & \mu < E_{c},
\\

C+A E_{c}+\frac{1}{2} B E_{c}^2 -A \mu-\frac{1}{2} B \mu^2 ,& & \mu
> E_{c}.
\end{array} \right. \label{50d}
\end{equation}
\end{widetext}

The variation $\delta N(E)$ of number of states looks as follows:
\begin{equation*}
\delta N(E)=\left \{ \begin{array}[c]{lcl} \frac{2}{3} \alpha
(E_{c}-E)^{3/2},& & E < E_{c},
\\
0 ,& &  E > E_{c}.
\end{array} \right.
\end{equation*}
Then the variation of the electron thermodynamic potential can be
written down in such a way

\begin{equation*} \delta \Omega(T,V,\mu) = -
\int\limits_{0}^{\infty}\frac{\delta N(\varepsilon)d \varepsilon
}{1+exp\left(\frac{\varepsilon-\mu}{T} \right)} \; .
\end{equation*}
We provide below the asymptotic expansion of the expression
$\delta \Omega(T,V,\mu)$  under condition $T \rightarrow 0,$
\begin{widetext}
\begin{equation}
\delta \Omega=\left \{ \begin{array}{lr} -\frac{4}{15}\alpha
E_{c}^{5/2} + \frac{4}{15}\alpha (E_{c}-\mu)^{5/2}
+\frac{\alpha}{6}\pi^2 T^2 (E_{c}-\mu)^{1/2}  +O(T^4), & \mu <
E_{c},
\\
-\frac{4}{15}\alpha E_{c}^{5/2}+\frac{\sqrt{\pi}}{2}\alpha T^{5/2}
\exp\left(\frac{-|E_{c}-\mu|}{T}\right)
erf\left(\sqrt{\frac{E_{c}}{T}}\right) +O\left( \left(
\exp\left(\frac{-|E_{c}-\mu|}{T}\right) \right)^2 \right), &
\mu>E_{c}. \label{51d}
\end{array} \right.
\end{equation}
\end{widetext} A similar expression was presented in the paper.
\cite{L}

\begin{figure}[!h]
\includegraphics[scale=0.65]{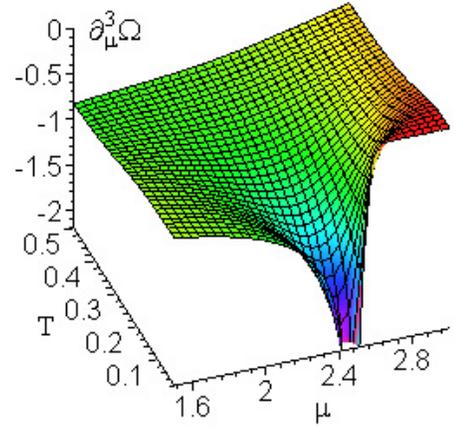}
\caption{\label{figure8} Third derivative of
thermodynamics potential, we put $V=20$, $\omega=0.3, \;
\omega\acute{}=0.5 i$ and $E_{c}=2.5$. Thermodynamics potential,
temperature and  chemical potential take on dimensionless values.
$\partial_{\mu}^{3} \Omega(T, V,\mu)$ has an infinite
 discontinuity when  $\mu \rightarrow
E_{c}, \: T=0$. }
\end{figure}

In Fig.~\ref{figure8}, the graph $\partial_{\mu}^{3} \Omega(T,
V,\mu)$ on the temperature $T$ and the chemical potential $\mu$ is
represented (while constructing, we used $V=20$, $\omega=0.3, \;
\omega'=0.5 i$, accordingly  $E_{c}=2.5$). One can see that when
$\mu \rightarrow E_{c}, \: T=0$ the third derivative from the
 thermodynamics potential has an infinite discontinuity and
\begin{equation*}
\lim \limits_{\mu \rightarrow E_{c}+0} \partial_{\mu}^{3}
\Omega(0, V,\mu)=0, \; \lim \limits_{\mu \rightarrow E_{c}-0}
\partial_{\mu}^{3} \Omega(0, V,\mu)=-\infty \; .
\end{equation*}

One can see also that the singularity at the point  $\mu=E_{c}$
smoothes out with  the temperature increase. In order to estimate an
appropriate temperature interval, we should pass from dimensionless
units to Kelvin degrees. In order to do this we have to multiply the
dimensionless temperature by the following coefficient,
\begin{equation}
\frac{E_{0}}{k_{b}}= \frac{\hbar^{2}}{2m a^{2} k_{b} } =
\frac{\hbar^{2}}{2m \left( c/2|\omega' \:| \right)^{2} k_{b} },
\label{52d}
\end{equation}
where $k_{b}$ is the Boltzmann constant  and $c$  is a period of
crystal potential. If  the potential period equals
$3\AA$ and effective electron mass is put $10m_{e}$, where $m_{e}$ - mass of free electron, we find out that the dimensionless temperature value $0.1$
corresponds to the temperature value in standard units $\sim
49^{0} K$. Since the width of the area $\bigtriangleup
(E_{c}-\mu), $ which contributes to the singularity of
$\partial_{\mu}^{3} \Omega(T, V,\mu),$  is proportional $T$, then
in a general case the singularity smoothes out slowly enough with
the temperature increase.

\section{Magnetization singularities of the Pauli paramagnetics} A
magnetization of an electron gas in magnetic field is described by
the following expression,

\begin{align}
M(T,\mu, H) & =\frac{\beta}{2} \int\limits_{0}^{\infty}\left( \frac{
1
}{1+exp\left(\frac{\varepsilon-\mu-\beta H}{T} \right)} \right. \nonumber \\
& \left. - \frac{ 1 }{1+exp\left(\frac{\varepsilon-\mu+\beta H}{T}
\right)} \right) \nu (\varepsilon) d\varepsilon, \label{53d}
\end{align}
where $\beta={e \hbar}/{2m c}$ is the Bohr magneton, $H$ is an
external magnetic field, $T$ is a temperature. According to this
expression the magnetization at $T=0$ can be presented in the
following way,
\begin{equation}
M(0,\mu, H)=M_{0} =\frac{\beta}{2} \left( N(\mu+\beta H) -
N(\mu-\beta H) \right) . \label{dd}
\end{equation}
Let us expand the expression (\ref{dd}) in the $(E_{c}-\mu\pm\beta
H)^{1/2}$ powers. Using the formula (\ref{49d}) we come to the
following result,
\begin{widetext}
\begin{equation*}
M_{0}= \left \{ \begin{array}[c]{lcl} \frac{\beta}{2}(2\beta H B
+\frac{2}{3}\alpha (E_{c}-\mu-\beta H)^{3/2}  -\frac{2}{3} \alpha
(E_{c}-\mu+\beta H)^{3/2} \\ \hspace{6cm} +O((E_{c}-\mu\pm\beta
H)^{5/2})),& & \mu < E_{c} - \beta H,
\\
\frac{\beta}{2}(2\beta H B - \frac{2}{3} \alpha (E_{c}-\mu+\beta
H)^{3/2} +O((E_{c}-\mu + \beta H)^{5/2}) ) ,& & E_{c} -\beta H < \mu
< E_{c} + \beta H,
\\
\beta^{2} H B,& & \mu > E_{c} + \beta H.
\end{array} \right.
\end{equation*}
\end{widetext}

Since  $\partial_{\mu} \nu (\mu \pm \beta H)$ has infinite
discontinuity  when $\mu \pm \beta H = E_{c}$ then the second
derivative of the appropriate magnetic moment with respect to the
chemical potential, $\partial_{\mu}^{2} M(0,\mu, H),$ has
singularity also when $\mu \pm \beta H = E_{c}$. The sign of
singularity is defined by the sign ahead of  $N(\mu \pm \beta H)$ in
the  formula for magnetization,
\begin{widetext}
\begin{equation*}
\partial_{\mu}^{2} M_{0}= \left \{ \begin{array}[c]{lcl}
\frac{\alpha\beta}{4(E_{c}-\mu-\beta H)^{1/2}}  -
\frac{\alpha\beta}{4(E_{c}-\mu+\beta H)^{1/2}} +O((E_{c}-\mu\pm\beta
H)^{1/2}),& & \mu < E_{c} - \beta H,
\\
- \frac{\alpha\beta}{4(E_{c}-\mu+\beta H)^{1/2}} +O((E_{c}-\mu +
\beta H)^{1/2}),& & E_{c} - \beta H < \mu  < E_{c} + \beta H,
\\
0,& & \mu > E_{c} + \beta H.
\end{array} \right.
\end{equation*}
\end{widetext}
The expression (\ref{d}) defines the relation  of the coefficient
$\alpha$ to the effective mass $m_{2}^{*}$. As mentioned above,
 with the decrease of gap width on  boundary of the Brillouin zone a value of the
coefficient $\alpha$ decreases. In the limit case when the gap
width equals zero the coefficient $\alpha =0$. In this case the
value $E_{c}$ is not critical point and there is no singularities
that is related to this point.

\begin{figure}[]
\includegraphics[scale=0.65]{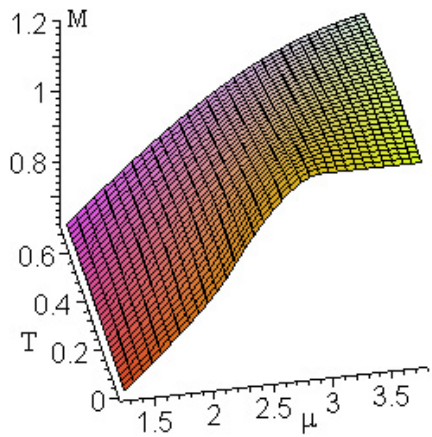}
\caption{\label{figure9} Magnetization of an electron
gas in magnetic field, we put $\beta H= 0.15 E_{c}$, $\omega=0.3, \;
\omega\acute{}=0.5 i$ and $E_{c}=2.5$. Temperature and chemical
potential take on dimensionless values, magnetization is measured in
Bohr magnetons.}
\end{figure}

In Fig.~\ref{figure9},  the graph $M(T,\mu, H)$ on the temperature
and the chemical potential is represented  (we put $\beta
H=0.15E_{c}$).

\begin{figure}[!]
\includegraphics[scale=0.65]{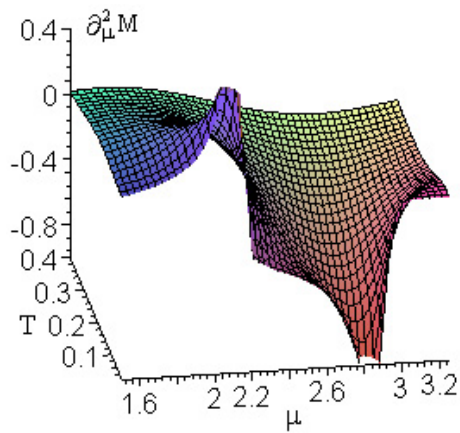}
\caption{\label{figure10} Second derivative of
electron gas magnetization in magnetic field, we put $\beta H= 0.15
E_{c}$, $\omega=0.3, \; \omega\acute{}=0.5 i$ and $E_{c}=2.5$.
Temperature and chemical potential take on dimensionless values,
magnetization is measured in Bohr magnetons. $\partial_{\mu}^{2}
M(T,\mu, H)$ has infinite discontinuities when  $\mu \rightarrow
E_{c} \pm \beta H, \: T=0$.}
\end{figure}

In Fig.~\ref{figure10}, the graph $\partial_{\mu}^{2} M(T,\mu, H)$
on the temperature and the chemical potential is represented (while
constructing, we used $\beta H= 0.15 E_{c}$, $\omega=0.3, \;
\omega'=0.5 i$, accordingly  $E_{c}=2.5$). One can see that when
$\mu \rightarrow E_{c} \pm \beta H$ at $T=0$ the second derivative
from the magnetic moment has infinite discontinuities and
\begin{equation*}
\lim \limits_{\mu \rightarrow E_{c}-\beta H-0} \partial_{\mu}^{2}
M_{0}=\infty, \; \lim \limits_{\mu \rightarrow E_{c}-\beta H+0}
\partial_{\mu}^{2} M_{0}=const ,
\end{equation*}
\begin{equation*}
\lim \limits_{\mu \rightarrow E_{c}+\beta H-0} \partial_{\mu}^{2}
M_{0}=-\infty, \; \lim \limits_{\mu \rightarrow E_{c}+\beta H+0}
\partial_{\mu}^{2} M_{0}=const .
\end{equation*}
As in the previous case, the singularities at the points $ E_{c} \pm
\beta H$ smooth out slowly enough when the temperature increases.

\section{Conclusions}

In this paper, we have examined the one-gap and two-gap separable
Lame potentials in detail. We have constructed the dispersion
relation $E(k)$ for one-dimensional case and the Fermi surfaces in
the first and second bands for three-dimensional case. The
pictures illustrate a passage from the limit case of free
electrons to the limit case of tight binding electrons. We have
provided also analytical expressions for effective masses of
electrons in a metal. These expressions depend on all the band
edges.

These results have been used to study the singularities of the
electron part of the thermodynamic characteristics in metals. We
have derived explicit analytical expressions for the density of
states in a metal and its derivative. Then we have examined the
thermodynamic potential and magnetic moment of metal in an external
magnetic field and extracted the singular part. We have also
obtained the relation between the parameter of singularity and
corresponding effective mass.  All the expressions we derived
contain  the parameters of our potential only. Therefore, if we fix
potential, we can calculate all the coefficients of these
expressions.

It is necessary to point out that instead of electrons, the
arbitrary quasi-particles can be considered within a framework of
this approach.

We would like also to say a few words about generalization of the
finite-gap potentials to lattices of other spacial symmetry. In
the paper \cite{BD} the authors have tried to develop such a
generalization for HCP lattice by means of the perturbation theory
and calculated the Fermi surface of Beryllium. The elliptic
Calogero-Moser potentials are also of significant interest
concerning their use in solid state physics. These potentials are
multidimensional.

\end{document}